# Unveiling the inter-layer interaction in a 1H/1T TaS$_2$ van de Waals heterostructure

*Cosme G. Ayani[1,2], M. Bosnar[3,4], F. Calleja[2], Andrés Pinar Solé[5], O. Stetsovych[5], Iván M. Ibarburu[1,2], Clara Rebanal[1,2], Manuela Garnica[2,6], Rodolfo Miranda[1,2,6,7], M. M. Otrokov[8], M. Ondráček[5], Pavel Jelínek[5], A. Arnau[3,4,9], Amadeo L. Vázquez de Parga[1,2,6,7]*

[1] *Departamento de Física de la Materia Condensada. Universidad Autónoma de Madrid, Cantoblanco, Madrid 28049, Spain.*

[2] *IMDEA Nanociencia. Calle Faraday 9, Cantoblanco, Madrid 28049, Spain.*

[3] *Departamento de Polímeros y Materiales Avanzados: Física, Química y Tecnología, Facultad de Ciencias Químicas, Universidad del País Vasco UPV/EHU, 20018 Donostia-San Sebastián, Spain.*

[4] *Donostia International Physics Center (DIPC). 20018 Donostia-San Sebastián, Spain.*

[5] *FZU - Institute of Physics of the Czech Academy of Sciences, Cukrovarnická. Cukrovarnicka 10, Prague 6, CZ 16200, Czech Republic.*

[6] *Instituto Nicolás Cabrera (INC). Universidad Autónoma de Madrid, Cantoblanco 28049, Madrid, Spain.*

[7] *Condensed Matter Physics center (IFIMAC), Universidad Autónoma de Madrid, Cantoblanco 28049, Madrid, Spain.*

[8] *Instituto de Nanociencia y Materiales de Aragón (INMA), CSIC-Universidad de Zaragoza, Zaragoza 50009, Spain*

[9] *Centro de Física de Materiales CSIC/UPV-EHU-Materials Physics Center, Manuel Lardizabal 5, E-20018 San Sebastián, Spain.*

ABSTRACT

This study delves into the intriguing properties of 1H/1T-TaS$_2$ van der Waals heterostructure, focusing on the transparency of the 1H layer to the Charge Density Wave of the underlying 1T layer. Despite the sizable interlayer separation and metallic nature of the 1H layer, positive bias voltages result in a pronounced superposition of the 1T charge density wave structure on the 1H layer. The conventional explanation relying on tunneling effects proves insufficient. Through a comprehensive investigation combining low-temperature scanning tunneling microscopy, scanning tunneling spectroscopy, non-contact atomic force microscopy, and first-principles calculations, we propose an alternative interpretation. The transparency effect arises from a weak yet substantial electronic coupling between the 1H and 1T layers, challenging prior understanding of the system. Our results highlight the critical role played by interlayer electronic interactions in van der Waals heterostructures to determine the final ground states of the systems.

INTRODUCTION

In recent years many two-dimensional (2D) layers of transition metal dichalcogenides (TMDs) have been synthetized owing to their interesting electronic structure, making them attractive from both fundamental and applied points of view. Furthermore, the relative simplicity with which these 2D materials can be combined into different van der Waals (vdW) heterostructures- by, e.g., the epitaxial growth or the exfoliation-and-stacking procedure, additionally enhances their versatility [1–3]. In particular, vdW heterostructures of TMDs offer an ideal playground for studying strongly correlated phases and phase transitions. One of the most fruitful examples of such heterostructures are stacks of different polymorphs of TaS$_2$ or TaSe$_2$, which can host several correlated ground states, like Kondo lattice, magnetic order and even a gapless spin liquid ground state, mimicking the physics already studied in heavy fermion materials [4–7].

To fully understand these phenomena, it is necessary to examine many factors that can determine them. For example, a low-temperature scanning tunnelling microscopy (STM) study of a 4Hb-TaS$_2$ system recently found zero bias excitations at vortices and step edges, suggesting that this compound could be a topological nodal superconductor [8, 9]. This behavior of the 4Hb-TaS$_2$ system is believed to emerge from admixing of its two constituent components that feature contrasting properties, analogous to heavy fermion systems such as CeCu$_2$Si$_2$ [10]. Namely, 4Hb-TaS$_2$ consists of interwoven monolayers of 1T-TaS$_2$ and 1H-TaS$_2$ phases, the former features a charge density wave (CDW) reconstruction, the so called star-of-David (SoD) [11, 12], below 183 K leading to high electron localization at low temperatures in a Mott insulator phase [11–13], while the latter also features another CDW reconstruction below 81 K but remains metallic and has a superconducting (SC) transition at around 1 K [14–16]. However, we find that the interlayer interaction between TaS$_2$ layers is poorly understood from a fundamental point of view and, therefore, the explanation of many observations is somewhat unsatisfying and deserves being revised.

As an example, we can consider the so called "transparency effect", in which low temperature STM images of 1H layer on one or more 1T layers at bias voltages in the range 100-250 mV clearly shows the periodicity of the 1T CDW superposed on the 1H surface [13, 17–21]. The original explanation of this transparency effect assigns it to a purely tunnelling effect, wherein electrons tunnel from the tip to the unoccupied states of 1T layer through the 1H layer, assuming that the layers themselves are decoupled [19–21].

However, the tunnelling current falls exponentially with distance, so given the large distance from tip to 1T layer due to the thickness of 1H layer and vdW gap, the signal coming from 1T layer should be much smaller than the 1H contribution. Furthermore, recent works have reported a subtle transparency effect even at negative bias voltages [23]. Therefore, a pure tunnelling effect with decoupled 1H and 1T layers is an insufficient explanation and there must be a different mechanism for such a clear transparency effect.

In order to unveil the origin of the coupling between layers, we have studied the vertical $TaS_2$ heterostructure that features a 1H and 1T monolayers stacked, by means of low temperature STM, scanning tunnelling spectroscopy (STS), Kolibri non-contact atomic force microscopy (NC-AFM) and first-principles calculations. The analysis of our data leads us to an alternative explanation of the origin of the 1H layer transparency effect that accounts for the 1T signal strength: it is clearly produced by a weak hybridization between the 1H and the 1T layer underneath and, thus, it involves interaction between the layers of the vdW heterostructure [24]. We believe that this electronic coupling between layers must be considered when explaining all phenomena concerning $TaS_2$ systems, including unconventional superconductivity [8, 9].

RESULTS AND DISCUSSION

The 1H/1T van der Waals heterostructures are studied in a 2H-$TaS_2$ crystal cleaved in UHV and transferred directly into the STM at a base temperature of 1.2 K. As investigated with STM, the surface of the crystal presents wide and clean areas with the presence of single layer step-edges, like the one depicted in Fig.1 a). In this case, the step edge runs along the center of the STM image with an apparent height of 680±70 pm and marks the division between two different stacking sequences: a 1T/2H on the left and a 1H/1T/2H on the right. In panels b) and c) of Fig.1 we present the top view of both 1T and 1H phases, as well as the lateral view of the complete stacking sequence. Both polymorphic phases of $TaS_2$ co-exist on the surface of the 2H crystal, probably due to their small difference in formation energy [16]. Indeed, a polytype transition can be induced via the STM tip, as shown in section S1 of the supplementary information, where a section of a 2H polytype is successfully transformed via the current of the STM tip into the 1T polytype [16, 20, 25].

We have characterized the atomic and electronic properties of the two constituent parts of the heterostructure, the results are depicted in Fig.2. The first row is devoted to the 1H layer while the second shows the results regarding the 1T layer. Atomically resolved STM images of the 1H layer, Fig.2 a), reveal the presence of two large periodicities, one is the quasi-(3×3) CDW from the H-type layer and the second one is the (√13×√13) R13.9° CDW from the 1T layer below. The two periodicities are better distinguished and studied via the fast Fourier transform (FFT) of the STM image as portrayed in panel b) of Fig.2. All the results of FFT analysis are presented in table S2 of the supplementary information and, all of them are in good agreement with those reported in other 2H crystals [9, 26]. The CDW of the 1H phase is a quasi-(3×3), exactly $2.7a_0$ (see section S3 in SI).

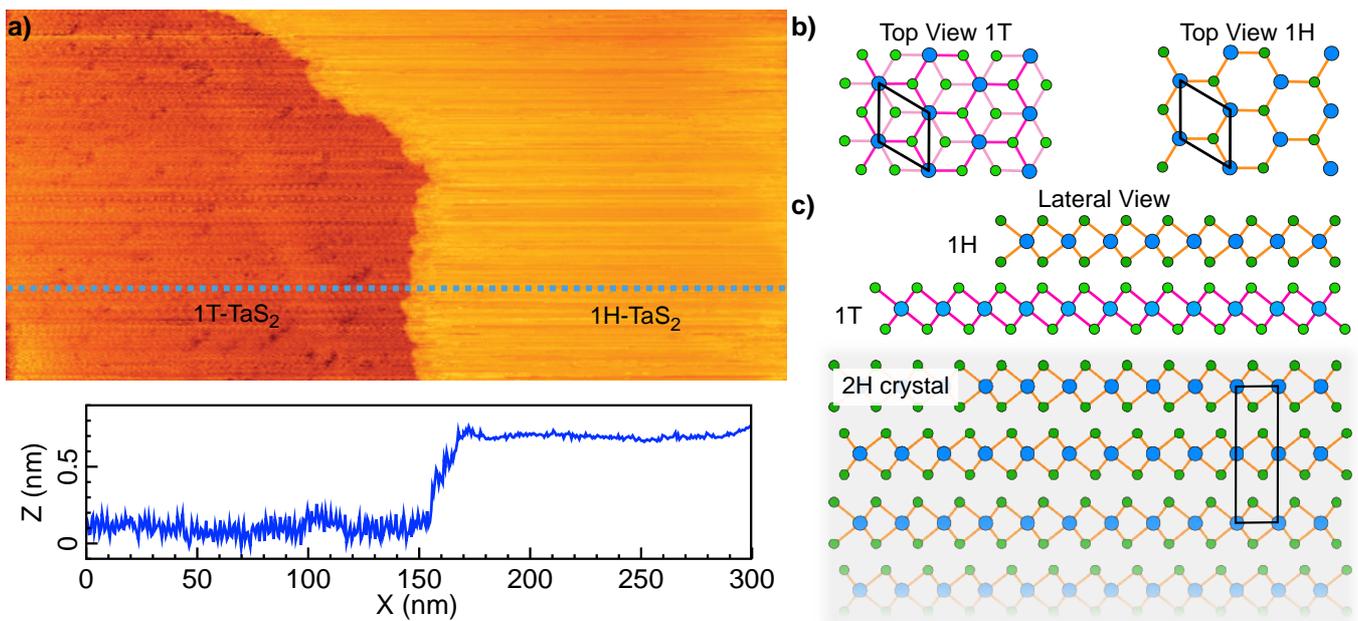

*Fig.1 1H/1T Heterostructure: a) 300 nm wide STM image of the 1H/1T heterostructure and corresponding line profile below. The 1H and 1T surfaces are separated by a single layer step-edge of 680± 70 pm. Image parameters: $V_b$=300 mV, I=100 pA. b) Ball models showing the top views of both the 1H and 1T phases, the blue balls represent the Tantalum atoms, the green balls represent the Sulphur atoms, and the orange and purple bonds represent the corresponding trigonal prismatic or octahedral coordination. c) Lateral view of the 1H/1T heterostructure on the*

surface of the 2H-TaS$_2$ crystal, the vertical unit cell is marked with a black rectangle in the case of the 2H crystal, notice the 180° rotation every two layers.

Figure 2 d) shows an atomically resolved image on the 1T layer, panel e) shows the corresponding FFT image. The rotation between the CDW and the atomic lattice is marked with the dashed purple and green line in the FFT. Both the atomic and CDW periodicities of the 1T layer of the heterostructure are in good agreement with those reported in 1T crystals [12, 14]. The atomic lattices of both layers are aligned as can be seen comparing Figure 2 b) and e) (see details in section S4 in SI).

Figure 2 c) shows the comparison between the dI/dVs measured on the 1H area (light blue) and the 2H crystal (dark blue). In both spectra the quasi-(3×3) CDW pseudo-gap appears at the Fermi level [23], [27], [28]. However, in the 1H areas we find a new feature between 230 ∼ 260 mV, which has not been reported on previous works on 4Hb-TaS$_2$ crystals [23] and, in the case of 1H layers embedded in 1T-TaS$_2$ crystal, it may be sightly distinguished in a spectrum at 4.2 K shown in the supplementary information of ref [16], but no explanation is given regarding its origin. Figure 2 f) shows the dI/dV measured on the 1T layer showing the presence of the Upper and Lower Hubbard band features [29, 30].

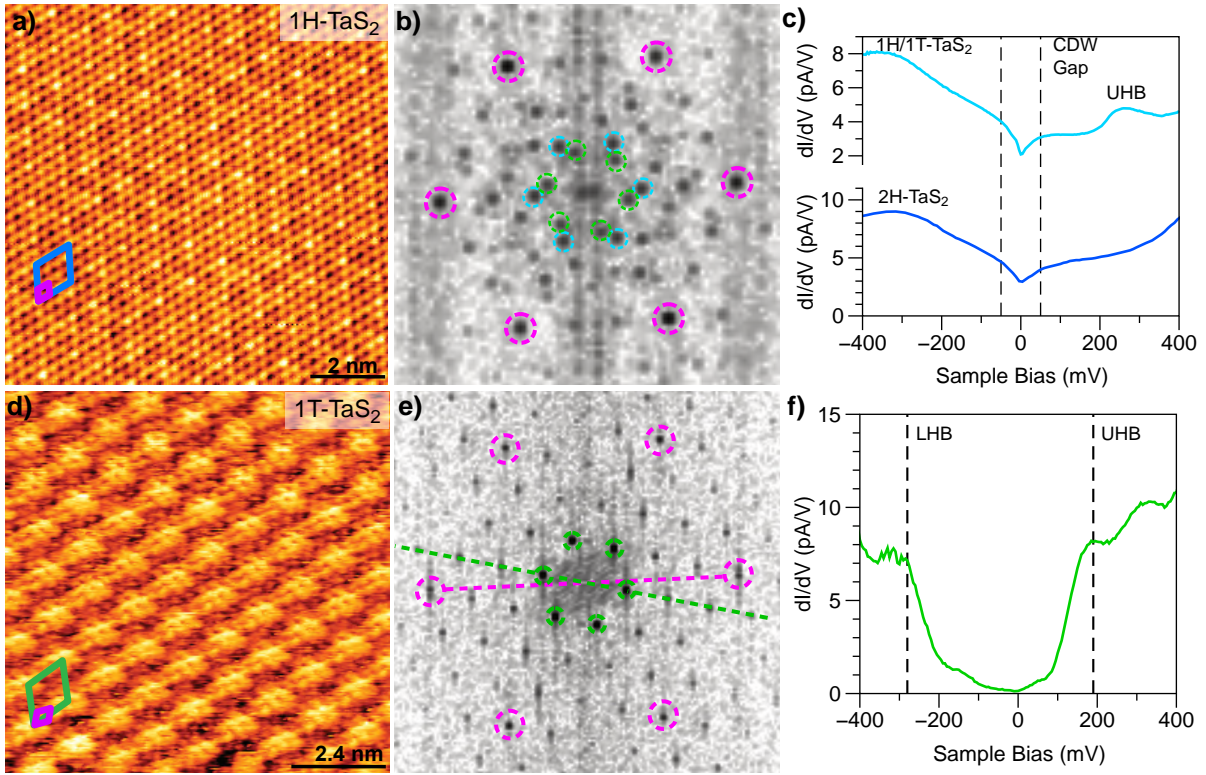

**Fig.2 1H/1T Heterostructure atomic and electronic properties**: *a) STM topography image of the 1H surface where two periodicities are marked with two rhombi, in purple for the atomic lattice and blue for the quasi-(3×3) CDW. Image parameters: $V_b$=100 mV, I=300 pA. b) FFT of the STM image in panel a), three periodicities are observed here and highlighted with circles. In purple the atomic lattice, in blue the quasi-(3×3) CDW and in green the (√13×√13)R13.9° CDW of the 1T layer underneath. c) STS point spectra of the 1H and 2H surfaces in light blue and dark blue, respectively. The main difference in the LDOS of the 1H layer is the hump at 260 mV marked as UHB. STS parameters: $V_b$=400 mV, I=500 pA, $V_{mod}$=4 mV. d) STM atomically resolved topography image of the 1T surface, the periodicities are marked with two rhombi, purple for the atomic lattice and green for the (√13×√13)R13.9° CDW. Image parameters: $V_b$=100 mV, I=300 pA. e) FFT of the 1T surface, the atomic and CDW periodicities are highlighted with purple and green circles, respectively. f) STS point spectrum of the 1T surface where the positions of the upper Hubbard band (UHB) and lower Hubbard band (LHB) are marked. STS parameters: $V_b$=500 mV, I=500 pA, $V_{mod}$=5 mV.*

The topography and FFT image of the 1H surface already reveals the presence of the CDW from the 1T layer underneath, in essence the images scanned over the 1H surface are bias dependent and the 1T CDW always appears superposed with varying intensity. This behavior suggests the presence of intricate electronic interactions between the 1H and 1T layers that are not immediately evident from direct tunneling considerations, as already argued. In order to unveil these interactions, we have studied the intensity of both CDW periodicities on the 1H surface for a wide range of positive bias voltages by means of FFT analysis of STM and AFM images, in order to track their bias voltage dependence. All the results are depicted in Fig.3 and Fig.4.

Figure 3 shows the results regarding STM data, in panel a) six STM images from a voltage series out of a total of twelve are depicted. The bias voltages at which the STM images are scanned are also marked on the STS spectrum of panel b), which was acquired on the same area and with the same tip used to complete the whole bias voltage series. Two spectral features are identified in the STS spectrum, one is the upper edge of the 1H layer conduction band that is resolved at 500 mV and the second is a prominent peak at +260 mV that roughly corresponds to the energy position of the UHB on the 1T phase underneath. The results of the bias voltage series are summarized in panel c). Here, the intensity of the 1st order FFT spots for the three different periodicities is plotted as a function of the bias voltage of the corresponding STM image. From both the STM images and the integrated FFT intensity plot, we can conclude that there is an inversion in the intensity of the CDWs at 100 mV below which the quasi-(3×3) dominates the topography of the 1H surface. For higher bias voltages the relation is the opposite, the superimposed (√13×√13)R13.9° becomes more intense following the same trend described in previous works [19–22].

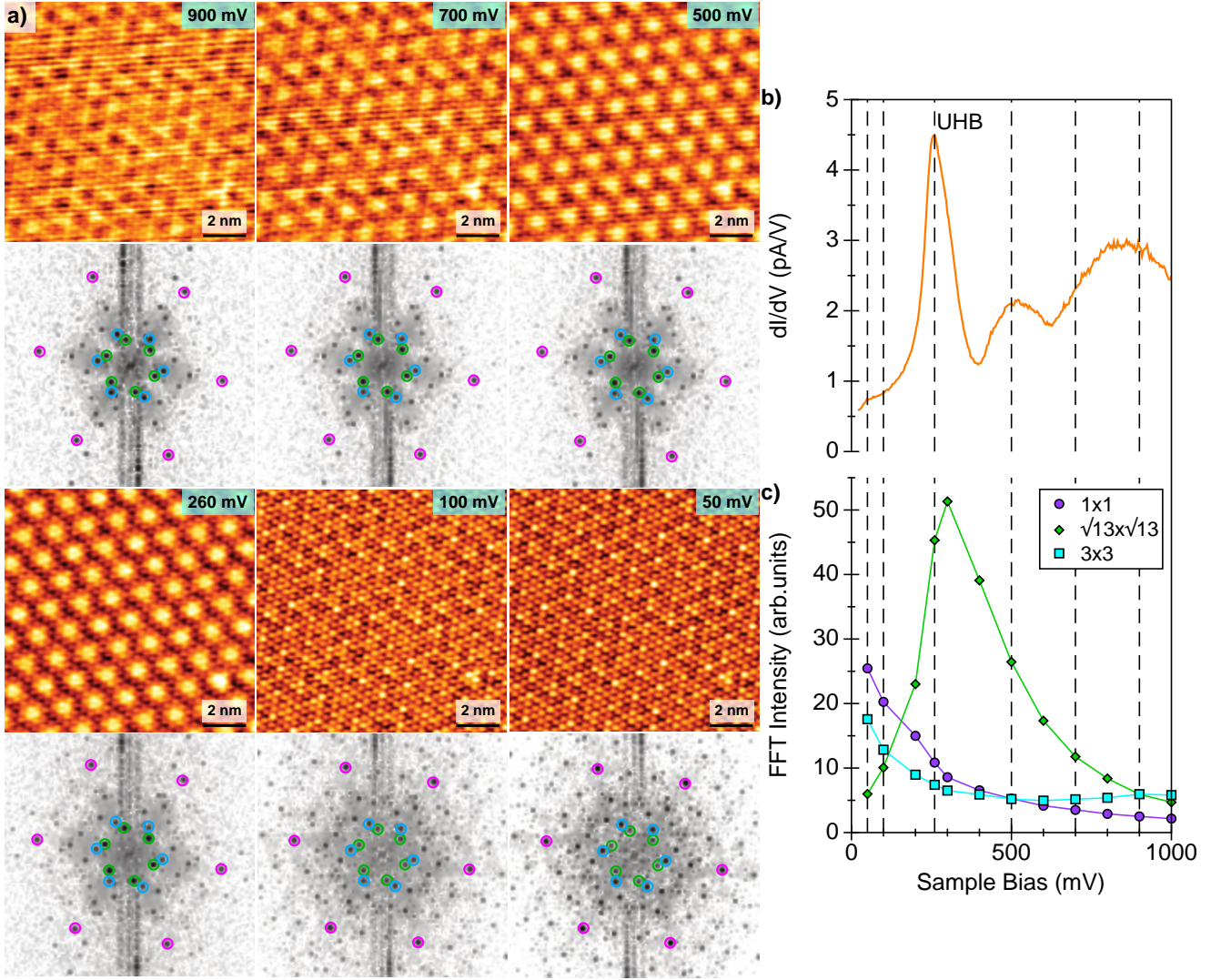

*Fig.3 Transparency effect in STM: a) STM Images and corresponding FFT plots performed on the 1H surface for different values of the bias voltage ($V_b$), the voltage value of the image is shown on the top right corner of the images. All STM images are scanned at a current of 300 pA. b) dI/dV spectrum performed on the same 1H area with the same tip used to perform the bias voltage series images shown in panel a). The bias voltage at which the images are taken is marked in the spectrum with black dashed lines, the hump observed at 500 mV is produced by the upper edge of the lowest 1H conduction band. The feature measured at +260 mV is related with the Upper Hubbard band from the 1T layer [29]. STS parameters: $V_b$=1 V, I=700 pA, $V_{mod}$ = 4 mV. c) Plot of the absolute intensity of the 1st order FFT spots for each periodicity as a function of the sample bias voltage.*

However, some differences and new information can be extracted from the FFT intensity plot in Fig.3 c). Above 900 mV, the intensities of both superstructures are practically the same, the intensity of the quasi-(3×3) being slightly higher for 1 V. As the voltage is decreased the intensity of the (√13×√13)R13.9° CDW spots increases reaching a maximum at 300 mV and then falls rapidly. Comparing panels b) and c) of Fig.3, we can see that the maximum intensity of the (√13×√13)R13.9° reconstruction is reached precisely in the vicinity of the +260 mV UHB feature and then decreases for lower voltages. To further confirm the relation between the STS feature and the 1T CDW periodicity, we have performed STS point spectra following the periodic unit cell of the (√13×√13)R13.9° CDW superposed on the 1H layer. The results, shown in Fig.S6 of the SI, indicate that the intensity maxima that appear in the dI/dV map at +260 mV correspond exactly to the 1T CDW maxima. Therefore, the +260 mV feature is spatially modulated with the 1T CDW that appears superimposed on the 1H layer. This effect is independent on the tip-sample distance (see sections S7 and S8 for detailed discussion).

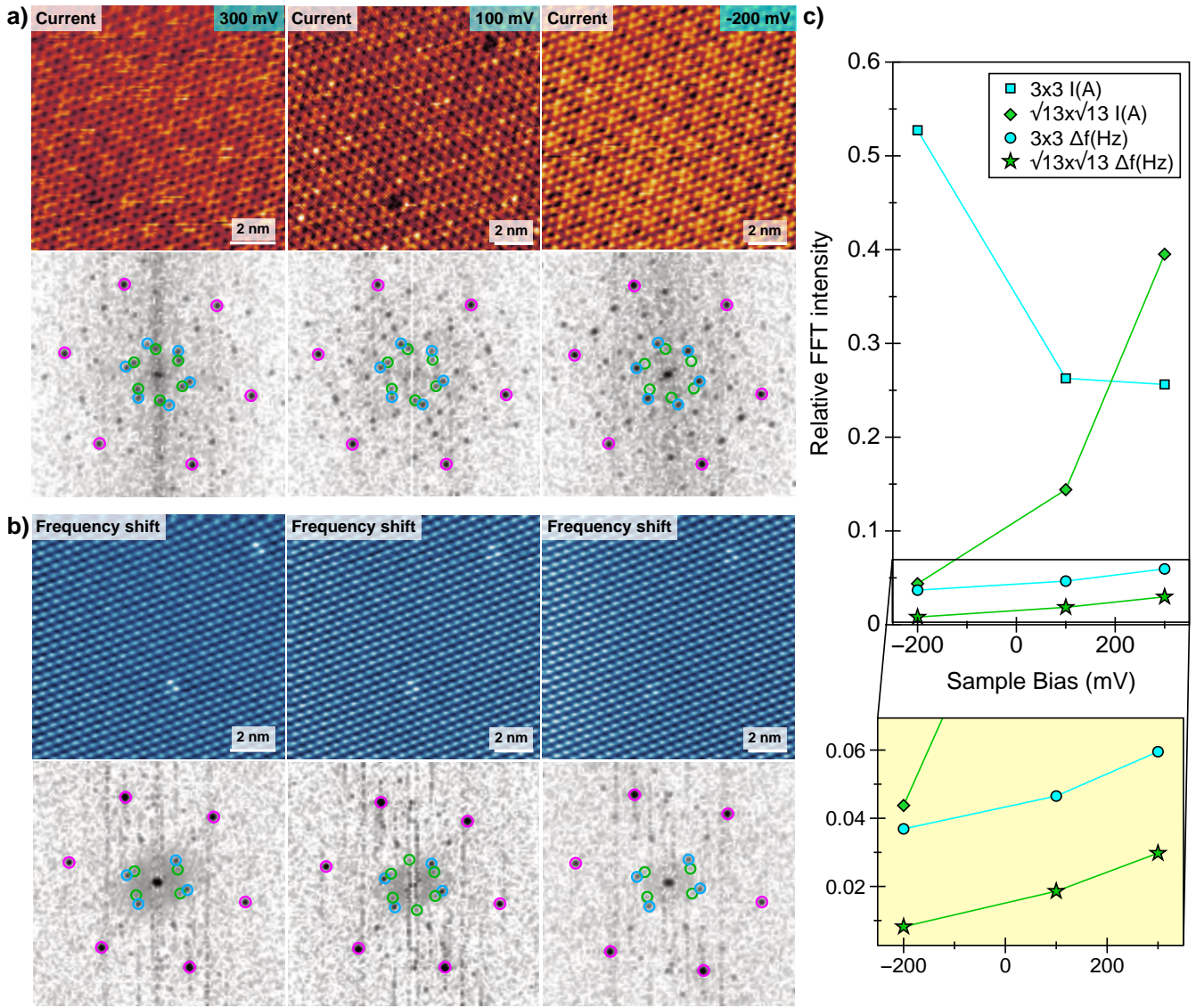

*Fig.4 Transparency effect in AFM: a)* AFM images performed with a Kolibri sensor on the 1H surface. In this panel the current channel is depicted, below each AFM image its corresponding FFT is shown where the 3 different periodicities are highlighted with circles. Only the voltage is varied between each image. *b)* Same AFM image as the one shown in panel a) but in this case the frequency shift channel is depicted, also the corresponding FFT image are shown below. *c)* The plot shows how the integrated intensity of the 1st order FFT spots for each periodicity of the current and frequency shift channel changes as the bias voltage is ramped. In this case the FFT intensity is normalized with respect to the intensity of the atomic periodicity FFT spots. The zoom below shows a closer look at the frequency shift results. Image parameters: I=10 pA, -1.15 Hz < Δf < 0.058 Hz.

Figure 4 a) shows the corresponding experimental results performed with an AFM, we show both the current channel, panel a), and the frequency shift channel, panel b). The results are condensed in the corresponding relative FFT intensity plot of panel c). In this case, due to the large signal from the atomic lattice, as compared to the STM experiments, the FFT intensity of both CDWs is normalized to the atomic lattice for clarity. Regarding the current channel, the behavior is analogous to the one observed in the STM, even the crossover in intensities is also measured at 100 mV, acting as control measurements and ensuring that the results of the frequency shift channel are reliable. Surprisingly, in the frequency channel we can resolve both CDWs as can be seen by the FFT analysis. However, unlike in the current channel, there is no crossover of intensities and the quasi-(3x3) CDW dominates over the 1H surface. Also, comparing with both current and STM measurements the relative intensity of the CDW spots with respect to the atomic lattice is much lower, representing always less than 6% of the atomic lattice intensity (see inset of Fig.4 c)). These results point in the direction that the transparency effect is mainly an electronic effect.

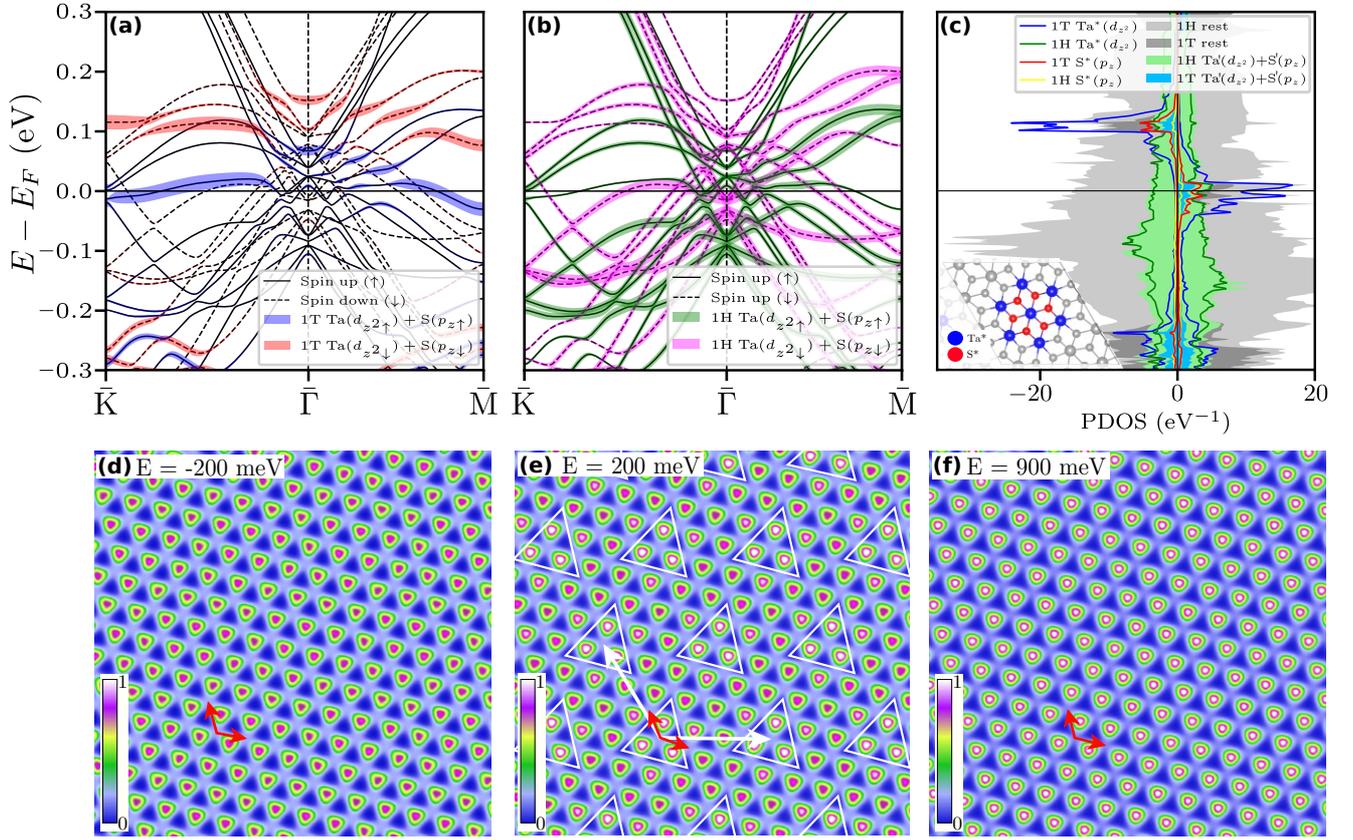

**Fig. 5 Explanation of the transparency effect**: *Electronic band structure of the 1H/1T bilayer with fat bands showing contributions of dominant out-of-plane orbitals Ta($d_{z^2}$) and S($p_z$) stemming from **a)** 1T and **b)** 1H layers. The PDOS analysis, panel **c)**, shows that the non-dispersive Hubbard bands are predominantly localized on the Ta atom at the center of the star-of-David distortion in the 1T layer and the closest Ta\* and S\* atoms (see the inset). The 1H Ta\* and S\* atoms are those closest to the 1T Ta\* and S\* atoms. The prime symbol [ ´] denotes all other Ta and S atoms in the heterostructure. Spin up and spin down channels appear in the right and left panels, respectively. The corresponding DOS contributions resonate with the contributions of out-of-plane orbitals of atoms in the 1H layer above them, resulting in a noticeable hybridization. Panels **d)-f)** show simulations of constant height integrated LDOS above the 1H layer. In panel (d) the energy integration interval does not contain the band crossings due to 1H-1T hybridization and the LDOS shows a (1×1) pattern (red arrows). In panel (e) this interval includes the band crossings and the LDOS instead shows a ($\sqrt{13}\times\sqrt{13}$) pattern (white arrows denote the periodicity). The white triangles are centered on Ta atoms at the center of the star-of-David distortion and serve as a guide to the eye. In panel (f) the energy range is increased further to 900 meV and the ($\sqrt{13}\times\sqrt{13}$) pattern is muted due to the small hybridization contribution compared to the total DOS (cf. the contribution of the rest of the system in panel (c)). The inset in panel (c) was produced with VESTA package [31], while the LDOS plane cuts were produced by XCrySDen package [32].*

In order to clarify the origin of this electronic effect we have performed state of the art DFT+U calculations. We consider as a minimal model the bilayer 1H-TaS$_2$/1T-TaS$_2$ heterostructure, i. e., the topmost two layers of the 1H/1T/2H heterostructure. The system is constructed following the approximation used in previous works [6, 33], where the lattice parameter of the ideal 1H layer in $\sqrt{13} \times \sqrt{13}$ supercell is extended 1% to match the CDW reconstructed 1T layer (see Methods section and Supplementary material for details).

The calculated electronic structure of the model 1H/1T bilayer is presented in Figure 5. By comparing the fat bands corresponding to the 1T Ta($d_{z^2}$) and S($p_z$) orbitals in the heterostructure (panel (a)) with the bands of freestanding 1T (see Supplementary material, Figure S9), the lower and upper Hubbard bands can be identified. The lower Hubbard band is partially occupied, forming two electronic pockets near the 2D Brillouin zone boundary points K and M, while the upper Hubbard band is found at 100 meV above the Fermi level. The fat bands corresponding to the same orbitals of 1H (panel (b)) show that the Hubbard bands also have a small amount of these 1H contributions, suggesting that hybridization is taking place. In fact, the projected density of states (PDOS, panel (c)) shows that the largest 1H contributions to the 1T Hubbard bands comes from the 1H atoms closest to the center of the star-of-David distortion.

This hybridization can be visually connected with the transparency effect using local density of states (LDOS) calculations. Figures 5(d)-(f) show the plane cut of the LDOS directly above the outer S atoms of the 1H layer integrated in different energy intervals. In the interval from E$_F$−200 meV to E$_F$ most of the bands have 1H character and, therefore, the corresponding LDOS (panel (d)) above the 1H layer has its 1x1 periodicity. In the interval from E$_F$ to E$_F$+200 meV, where the hybridization of 1T Hubbard bands with 1H bands was found to occur, the LDOS acquires the $\sqrt{13} \times \sqrt{13}$ periodicity instead (panel (e)). When the integration is extended to a larger interval, e.g., from E$_F$ to E$_F$ + 900 meV as in panel (f), the $\sqrt{13} \times \sqrt{13}$ periodicity in LDOS is again not visible because there are many additional 1H bands which are not hybridized with the 1T Hubbard bands. This behaviour of the LDOS agrees with the STM observation shown in Figure 3(a).

Note that, while the agreement of our model and the experimental results shows that this model can explain the origin of the transparency effect, one must be aware of its limitations. In particular, the energy position of the calculated Hubbard bands does not match the observed features in STS. For example, the UHB is observed at 260 mV in STS, while it appears around 100 meV in our calculations. Nevertheless, what matters is that, regardless the energy position of the Hubbard bands, the 1H and 1T hybridization should happen if (i) atoms are close enough for S($p_z$) orbitals in the 1H and 1T layers to have a sizeable overlap and (ii) 1H bands resonate with the states of the 1T Hubbard bands, for which there is a relatively wide energy window as seen in Fig. 3 a) and b).

## CONCLUSIONS

The use of low temperature STM and non-contact AFM has enabled us to characterize the geometric and electronic structure of 1H/1T TaS$_2$ heterostructure on a 2H crystal, finding the transparency effect of the 1H layer and an additional intriguing feature that appears in the 1H LDOS around + 260 mV. Our results show that there is a weak, although sizeable, coupling between the 1H and 1T layers. It is weak enough to maintain the structural properties of the vertical 1H/1T heterostructure with the same atomic lateral spatial periodicity, as well as the two CDWs periodicities of the pristine 1H and 1T layers, respectively, the quasi-(3x3) and ($\sqrt{13}\times\sqrt{13}$)R13.9° periodicities. However, this coupling between the 1T and 1H layers is strong enough to introduce hybridization between 1T and 1H electronic bands. A robust proof of this coupling is the observation of the 1T ($\sqrt{13}\times\sqrt{13}$)R13.9° periodicity on the 1H layer in STM images, happening for certain values of the applied bias voltage that fall close to the position of the UHB hybridized with the 1H bands.

In combination with the results of first-principles DFT calculations for a model 1H/1T TaS$_2$ bilayer, we have succeeded in giving a physical explanation to this, so called, transparency effect that is beyond a pure tunneling effect simply based on the uncoupled electronic densities of the 1T and 1H layers. Our electronic band structure and projected density of states analysis shows that hybridization between bands with 1T and 1H character near the Fermi level appears at relatively large van der Waals interlayer distances but close enough to have overlap between S($p_z$) orbitals in the 1H and 1T layers. In particular, the hybridization of the non-dispersive pristine upper Hubbard band in the 1T layer with ($\sqrt{13}\times\sqrt{13}$)R13.9° CDW with unoccupied bands near the Fermi level in the 1H layer translates into local density of states modulation in the 1H layer with this periodicity, as observed in the STM images.

This case study emphasizes the significance of interlayer interactions in a specific TaS$_2$ van der Waals heterostructure, with potential applicability to analogous systems where these weak interactions are cornerstone to understanding the reported low-temperature ground states, such as unconventional superconductivity.

## METHODS

### STM experiments
All experiments have been carried out in a UHV chamber with a base pressure of $5 \times 10^{-11}$ mbar equipped with a commercial Joule Thompson STM (JT-STM) from SPECS GmbH and facilities for tip and sample preparation. The STM is fitted with a superconducting coil that can provide a magnetic field up to 3 T perpendicular to the plane of the sample and therefore parallel to the tip axis. All STM/STS data were measured with tip and sample thermalized at 1.2 K.

### AFM experiments
The AFM experiments were carried out with a commercial low-temperature (1.2 K) STM/nc-AFM microscope (Specs-JT Kolibri sensor: f ≈ 1 MHz). The CO high resolution images were performed at constant height mode, 50 pm amplitude setpoint after functionalizing the Kolibri W tip with a CO from the bare Au(111) surface. The bias voltage was changed as detailed in the main text.

### Sample preparation
The 2H-TaS$_2$ crystal is adhered to the sample holder with RS pro silver conductive epoxy, the sample is then mounted on a cleaving set-up in the load-lock chamber and cleaved in UHV conditions. Also close to the sample a small gold ball is adhered with the same epoxy to perform in-situ tip indentations. In this way the tip may be prepared without having to take the sample out of the microscope and having to repeat the cleaving process.

### DFT calculations
The DFT calculations were performed using the Vienna Ab-initio Simulation Package (VASP) [34–36]. The lateral cell parameter was optimized for both the 3 × 3 1H-TaS$_2$ (1H) and $\sqrt{13} \times \sqrt{13}$ 1T-TaS$_2$ (1T) supercells in both the ideal and CDW distorted atomic configurations by total energy calculations of systems with varying lateral cell parameters and no spin polarization. These calculations were carried out with the VASP's PAW PBE datasets [37], a plane wave cutoff value of 275 eV, 7 × 7 × 1 Monkhorst-Pack sampling of the Irreducible Brillouin zone (IBZ), Gaussian smearing of the electronic occupations with the smearing factor of 10 meV and the electronic convergence threshold was $10^{-6}$ eV. The exchange-correlation functional was that of Perdew, Burke and Ernzerhof (PBE) [38] and the van der Waals interaction was taken into account through Grimme D3 model with Becke-Johnson cutoff [39, 40]. VASP's conjugate gradient algorithm was employed for the relaxation of atomic positions and the process was stopped when the force per atom decreased under 0.01 eV/Å.

In all subsequent calculations the spin polarization was taken into account and the GGA+U formalism [41] within the Dudarev scheme [42] was applied. The U and J parameters of $U_{1T}$ = 2.13 eV, $J_{1T}$ = 0.37 eV, $U_{1H}$ = 3.18 eV and $J_{1H}$ = 0.36 eV were used. These numbers were estimated by constrained random phase approximation (cRPA) calculations [43–45] as implemented in VASP. The Wannier functions required in cRPA calculations were obtained with the Wannier90 code [46, 47] and the polarization matrix was constructed according to Sasioglu-Friedrich-Blügel method [48]. See Supplementary material for more details regarding the cRPA calculations.

To obtain the band structure, projected (P) and local (L) densites of states (DOS) of the 1H/1T bilayer heterostructure, first a static calculation was performed with the same parameters as for relaxations and optimizations but with the Blöchl tetrahedron integration method instead of a Gaussian smearing of electronic occupations. Afterwards, the PDOS and LDOS were calculated by a non-self-consistent static calculation where the IBZ was sampled by a 15 × 15×1 Monkhorst-Pack grid, while bands were calculated in a similar manner along the k-point path between K-Γ-M high-symmetry points (again with Gaussian smearing of 10 meV).

For the sake of comparison, the band structure and projected density of states of the model 1H-1T TaS$_2$ bilayer heterostructure were also calculated with the Fritz Haber Institute Ab Initio Materials Simulations (FHI-AIMS) code [49]. The FHI-AIMS calculations were done on the same geometry that was used for VASP calculations. These calculations were done in the generalized gradient approximation (GGA) for the exchange-correlation functional, specifically in the PBE parametrization. Spin polarization was considered. The "light" default atomic basis sets for Ta and S as provided with the code was used. The 2D Brillouin zone was sampled by 8×8×1 k-points in the electron self-consistency cycle, while a 24 × 24 × 1 k-mesh and Gaussian smearing with σ = 0.05 eV was applied in the post-processing in order to obtain the layer-projected densities of states. A good agreement between the results obtained using FHI-AIMS and VASP was found, see Suppl. Fig. S11.

# Acknowledgments

This work has been supported by the Spanish Ministry of Science and Innovation, Grants no. PID2021-123776NBC21 (CONPHASETM), PID2019-103910GB-I00, PID2022-138210NB-I00 and PID2021-128011NB-I00 from MICIN/AEI/10.13039/501100011033 "ERDF A way of making Europe", also through "Severo Ochoa" (Grant CEX2020-001039-S) and "María de Maeztu" (Grant CEX2018-000805-M) Programmes for Centres of Excellence in R&D, respectively. Additionally, computational resources were provided by the e-INFRA CZ project (ID:90254), supported by the Ministry of Education, Youth and Sports of the Czech Republic. Financial support has also been recieved by the Comunidad de Madrid (Project S2018/NMT-4511, NMAT2D) and the Basque Government IT-1527-22. Financial support through the (MAD2D-CM)- MRR MATERIALES AVANZADOS-IMDEA-NC and (MAD2D-CM) MRR MATERALES AVANZADOS-UAM is also acknowledged. M.G. has received financial support through the "Ramón y Cajal" Fellowship program (RYC2020-029317-I) and "Ayudas para Incentivar la Consolidación Investigadora" (CNS2022-135175). M. O. and P. J. acknowledge the financial support from GACR project 20-13692X.

# Table of Contents

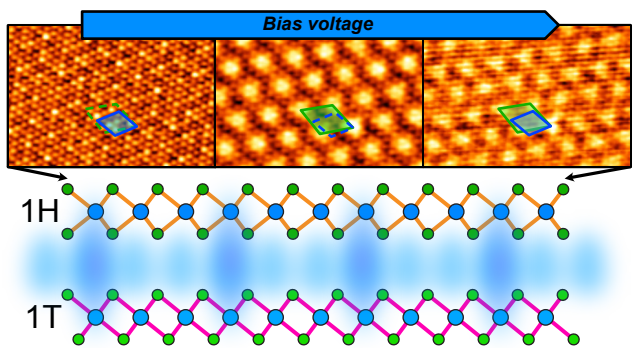

BIBLIOGRAPHY

1. M. Chhowalla et al., "The chemistry of two-dimensional layered transition metal dichalcogenide nanosheets," Nature_Chemistry, vol. 5, no. 4, pp. 263–275, 2013, doi: 10.1038/nchem.1589.
2. S. Manzeli et al., "2D transition metal dichalcogenides," Nature Reviews Materials, vol. 2, no. 8, pp. 17033, 2017, doi:_10.1038/natrevmats.2017.33.
3. Q. H. Wang et al., "Electronics and optoelectronics of two-dimensional transition metal dichalcogenides," Nature_Nanotechnology, vol. 7, no. 11, pp. 699–712, 2012, doi: 10.1038/nnano.2012.193.
4. C. G. Ayani et al., "Probing phase transition to a coherent 2D Kondo lattice," Small, vol., no., pp. 2303275, 2023,_doi:10.1002/smll.202303275.
5. V. Vaňo et al., "Artificial heavy fermions in a van der Waals heterostructure," Nature, vol. 599, no. 7886, pp. 582–586, 2021,_doi:10.1038/s41586-021-04021-0.
6. W. Wan et al., "Evidence for ground state coherence in a two-dimensional Kondo lattice," Nature communications, vol. 14,_no., pp.7005, 2023, doi: arXiv:10.1038/s41467-_023-42803-4.
7. W. Ruan et al., "Evidence for quantum spin liquid behaviour in single-layer 1T-TaSe2 from scanning tunnelling microscopy," Nature Physics, vol. 17, no. 10, pp. 1154–1161, 2021, doi: 10.1038/s41567-021-01321-0.
8. A. Ribak et al., "Chiral superconductivity in the alternate stacking compound 4Hb-TaS2," Science Advances, vol. 6, no. 13, p.eaax9480, 2020, doi: 10.1126/sciadv.aax9480.
9. A. K. Nayak et al., "Evidence of topological boundary modes with topological nodal-point superconductivity," Nature Physics, vol.17, no. 12, pp. 1413–1419, 2021, doi: 10.1038/s41567-021-01376-z.
10. W. Lieke et al., "Superconductivity in CeCu2Si2," Journal of Applied Physics, vol. 53, no. 3, pp. 2111–2116, 1982, doi:10.1063/1.330714.
11. P. Fazekas et al., "Electrical, structural and magnetic properties of pure and doped 1T-TaS$_2$," Philosophical Magazine B, vol. 39, no. 3, pp. 229–244, 1979, doi: 10.1080/13642817908245359.
12. J. A. Wilson et al, "The transition metal dichalcogenides discussion and interpretation of the observed optical, electrical and structural properties," Advances in Physics, vol. 18, no. 73, pp. 193–335, 1969, doi: 10.1080/00018736900101307.
13. R. V. Coleman et al., "Scanning tunnelling microscopy of charge-density waves in transition metal chalcogenides," Advances in Physics, vol. 37, no. 6, pp. 559-644, 1988. doi: 10.1080/00018738800101439.
14. J. A. Wilson et al., "Charge-density waves and superlattices in the metallic layered transition metal dichalcogenides," Advances in Physics, vol. 24, no. 2, pp. 117–201, 1975, doi: 10.1080/00018737500101391.
15. H. M. Lefcochilos-Fogelquist et al., "Substrate-induced suppression of charge density wave phase in monolayer 1H-TaS2 on Au(111)," Physical Review B, vol. 99, no. 17, pp. 1–6, 2019, doi: 10.1103/PhysRevB.99.174113.
16. Z. Wang et al., "Surface-Limited Superconducting Phase Transition on 1 T-TaS2," ACS Nano, vol. 12, no. 12, pp. 12619–12628, 2018, doi: 10.1021/acsnano.8b07379.
17. B. Giambattista et al.,"Corrrelation of scanning-tunneling-microscope image profiles and charge-density-wave amplitudes," Physical Review B, vol. 38, no. 5, pp. 3545, 1988, doi:10.1103/PhysRevB.38.3545.
18. Y. Fujisawa et al., "Superposition of (√13×√13) and (3×3) supermodulations in TaS2 probed by scanning tunneling microscopy," Journal of Physics: conference series, vol. 969, no. 012053, pp. 147-6596, 2018, doi: 10.1088/1742-6596/969/1/012053.
19. W. Han et al., "Bias-dependent STM images of charge-density waves on TaS2," Physical Review B, vol. 50, no. 19, pp. 14746–14749, 1994, doi: 10.1103/PhysRevB.50.14746.
20. J. J. Kim et al., "Atomic- and electronic-structure study on the layers of 4Hb-TaS2 prepared by a layer-by-layer etching technique," Physical Review B, vol. 52, no. 20, pp. R14388–R14391, 1995, doi: 10.1103/PhysRevB.52.R14388.
21. I. Ekvall et al., "Atomic and electronic structures of the two different layers in 4Hb at 4.2 K," Physical Review B, vol. 55, no. 11, pp. 6758–6761, 1997, doi: 10.1103/PhysRevB.55.6758.
22. C. Wen et al., "Roles of the narrow electronic band near Fermi level in 1T-TaS2 related layered materials," Physical Review_Letters, vol. 126, no. 25, pp. 256402-1, 2021, doi: 10.1103/PhysRevLett.126.256402.
23. S. Shen et al., "Coexistence of Quasi-two-dimensional Superconductivity and Tunable Kondo Lattice in a van der Waals Superconductor," Chinese Physical Letters, vol. 39, no. 7, pp. 077401, 2022, doi: 10.1088/0256-307X/39/7/077401.
24. B. Shao et al., "Engineering Interlayer Hybridization in Energy Space via Dipolar Overlayers," Chinese Physical Letters, vol. 40, no. 8, pp. 087303, 2023, doi: 10.1088/0256-307X/40/8/087303.

25. J. Zhang et al., "Creation of Nanocrystals Through a Solid-Solid Phase Transition Induced by an STM Tip," Science, vol._274, no. 5288, pp. 757–760, 1996, doi: 10.1126/science.274.5288.757.
26. Y. Yang et al., "Enhanced superconductivity upon weakening of charge density wave transport in 2H-TaS2 in the two-dimensional limit," Physical Review B, vol. 98, no. 3, pp. 035203-9, 2018, doi: 10.1103/PhysRevB.98.035203.
27. J. Hall et al., "Environmental Control of Charge Density Wave Order in Monolayer 2H-TaS2," ACS Nano, vol. 13, no. 9, pp. 10210–10220, 2019, doi: 10.1021/acsnano.9b03419.
28. C. Wang et al.,"Spectroscopy of dichalcogenides and trichalcogenides using scanning tunneling microscopy," Journal of Vacuum Science & Technology B: Microelectronics and Nanometer Structures, vol. 9, no. 2, pp. 1048,1991, doi:10.1116/1.585257.
29. S. Qiao et al., "Mottness collapse in 1T-TaS2-xSex transition-metal dichalcogenide: An interplay between localized and itinerant orbitals," Physical Review X, vol. 7, no. 4, pp. 041054-10, 2017, doi: 10.1103/PhysRevX.7.041054.
30. J.-J. Kim et al., "Observation of Mott Localization Gap Using Low Temperature Scanning Tunneling Spectroscopy in Commensurate 1T−TaS2," Physical Review Letters, vol. 73, no. 15, pp. 2103–2106, 1994, doi: 10.1103/PhysRevLett.73.2103.
31. K. Momma et al., "ıt VESTA3 for three-dimensional visualization of crystal, volumetric and morphology data," Journal of Applied Crystallography, vol. 44, no. 6, pp. 1272–1276, 2011, doi: 10.1107/S0021889811038970.
32. A. Kokalj, "XCrySDen - a new program for displaying crystalline structures and electron densities," Journal of Molecular Graphics and Modelling, vol. 17, no. 3, pp. 176-179, 1999, doi:10.1016/S1093-3263(99)00028-5.
33. A. K. Nayak et al., "First Order Quantum Phase Transition in the Hybrid Metal-Mott Insulator Transition Metal Dichalcogenide 4Hb-TaS2", arXiv:2303.01447, 2023, doi:10.48550/arXiv.2303.01447.
34. G. Kresse et al., "Efficient iterative schemes for ab initio total-energy calculations using a plane-wave basis set," Physical Review B, vol. 54, no. 16, pp. 11169–11186,1996, doi: 10.1103/PhysRevB.54.11169.
35. G. Kresse et al., "Ab initio molecular dynamics for liquid metals," Physical Review B, vol. 47, no. 1, pp. 558–561, 1993, doi:10.1103/PhysRevB.47.558.
36. G. Kresse et al., "Ab initio molecular-dynamics simulation of the liquid-metal–amorphous-semiconductor transition in germanium," Physical Review B, vol. 49, no. 20, pp. 14251–14269, 1994, doi: 10.1103/PhysRevB.49.14251.
37. G. Kresse et al., "From ultrasoft pseudopotentials to the projector augmented-wave method," Physical Review B, vol. 59, no. 3, pp. 1758–1775, 1999, doi: 10.1103/PhysRevB.59.1758.
38. J. P. Perdew et al., "Generalized Gradient Approximation Made Simple," Physical Review Letters, vol. 77, no. 18, pp. 3865–3868, 1996, doi: 10.1103/PhysRevLett.77.3865.
39. S. Grimme et al., "A consistent and accurate *ab initio* parametrization of density functional dispersion correction (DFT-D) for the 94 elements H-Pu," The journal of chemical physics, vol. 132, no. 15, pp. 154104, 2010, doi: 10.1063/1.3382344.
40. S. Grimme et al., "Effect of the damping function in dispersion corrected density functional theory," Journal of computational chemistry, vol. 32, no. 7, pp. 1456-1465, 2011, doi: 10.1002/jcc.21759.
41. V. I. Anisimov et al., "Band theory and Mott insulators: Hubbard U instead of Stoner I," Phyical Review B, vol. 44, no. 3, pp. 943–954, 1991, doi: 10.1103/PhysRevB.44.943.
42. S. L. Dudarev et al., "Electron-energy-loss spectra and the structural stability of nickel oxide: An LSDA+U study," Physical Review B, vol. 57, no. 3, pp. 1505–1509, 1998, doi: 10.1103/PhysRevB.57.1505.
43. L. Vaugier et al., "Hubbard U and Hund exchange J in transition metal oxides: Screening versus localization trends from constrained random phase approximation," Physical Review B, vol. 86, no. 16, pp. 165105, 2012, doi:10.1103/PhysRevB.86.165105.
44. T. Miyake et al., "Ab initio procedure for constructing effective models of correlated materials with entangled band structure," Physical Review B, vol. 80, no. 15, pp. 155134, 2009, doi: 10.1103/PhysRevB.80.155134.
45. T. Miyake et al., "Screened Coulomb interaction in the maximally localized Wannier basis," Physical Review B, vol. 77, no. 8, pp. 085122, 2008, doi: 10.1103/PhysRevB.77.085122.
46. A. A. Mostofi et al., "An updated version of Wannier90: A tool for obtaining maximally-localised Wannier functions," Computer Physics Communication., vol. 185, no. 8, pp. 2309–2310, 2014, doi: https://doi.org/10.1016/j.cpc.2014.05.003.
47. G. Pizzi et al., "Wannier90 as a community code: new features and applications," Journal Condensed Matter Physics, vol. 32, no. 16, pp. 165902, 2020, doi: 10.1088/1361-648x/ab51ff.
48. E. Sasioglu et al., "Effective Coulomb interaction in transition metals from constrained random-phase approximation," Physical Review B, vol. 83, no. 12, pp. 121101, 2011, doi: 10.1103/PhysRevB.83.121101.
49. V. Blum et al., "Ab initio molecular simulations with numeric atom-centered orbitals," Computer Physics Communications, vol. 180, no. 11, pp. 2175–2196, 2009, doi: https://doi.org/10.1016/j.cpc.2009.06.022.